\pdfoutput=1
\documentclass[twocolumn]{article}

\usepackage{tabulary,graphicx,times,caption,fancyhdr,amsfonts,amssymb,amsbsy,latexsym,amsmath}
\usepackage[utf8]{inputenc}
\usepackage{url,multirow,morefloats,floatflt,cancel,tfrupee,textcomp,colortbl,xcolor,pifont}
\usepackage[nointegrals]{wasysym}
\makeatletter

\usepackage{ifxetex}
\ifxetex\else
  \usepackage{dblfloatfix}
\fi

\@ifundefined{subparagraph}{
\def\subparagraph{\@startsection{paragraph}{5}{2\parindent}{0ex plus 0.1ex minus 0.1ex}%
{0ex}{\normalfont\small\itshape}}%
}{}

\def\URL#1#2{\@ifundefined{href}{#2}{\href{#1}{#2}}}

\def\UrlOrds{\do\*\do\-\do\~\do\'\do\"\do\-}%
\g@addto@macro{\UrlBreaks}{\UrlOrds}

\makeatother

\usepackage[paperheight=11in,paperwidth=8.3in,margin=2.5cm,headsep=.7cm,top=2.5cm]{geometry}
\usepackage[T1]{fontenc}

\renewenvironment{abstract}
	{\trivlist\item[]\leftskip0pt\par\vskip4pt\noindent
  	\textbf{\abstractname}\mbox{\null}\\}
	{\par\noindent\endtrivlist}

\def\keywords#1{\par\medskip\par\noindent\textbf{Keywords}: #1\par}

 \date{} \emergencystretch 8pt

\makeatletter
\def\author#1{\gdef\@author{\hskip-\tabcolsep%
	\parbox{\textwidth}{\raggedright\bfseries#1\\[1pc]}}}
\def\address[#1]#2{\g@addto@macro\@author{\\\hskip-\tabcolsep\parbox{\textwidth}{\raggedright%
	\normalsize\normalfont\textsuperscript{#1}#2}}}
\let\addresslink\textsuperscript
\def\correspondence#1{\g@addto@macro\@author{\\\hskip-\tabcolsep\parbox{\textwidth}{\raggedright%
	\vspace*{10pt}\normalsize\normalfont~\\#1~\\[12pt]}}}
\def\email#1{\g@addto@macro\@author{\\\hskip-\tabcolsep\parbox{\textwidth}{\raggedright%
	\normalsize\normalfont Emails: #1}}}

\def\title#1{\gdef\@title{\vspace*{-30pt}%
	\raggedright\textbf{\@journaltitle}~\\%
  \raggedright\bfseries\ifx\@articleType\@empty\vspace*{20pt}\else%
  \vspace*{20pt}\@articleType\vspace*{20pt}\\\fi#1}}
\let\@journaltitle\@empty \def\journaltitle#1{\gdef\@journaltitle{{\normalfont\itshape#1}}}
\let\@articleType\@empty \def\articletype#1{\gdef\@articleType{{\normalfont\itshape#1}}}

\let\@runningHead\@empty \def\RunningHead#1{\gdef\@runningHead{{\normalfont #1}}}

\usepackage{fancyhdr}
\fancypagestyle{headings}{\fancyhf{}
  \fancyhead[R]{\itshape\@runningHead}
  \fancyfoot[C]{\thepage}}
\usepackage{float,xcolor}

\usepackage{tabularx}
\usepackage{biblatex}
\newcounter{protocol}

\newcommand{\sbline}{\\[.5\normalbaselineskip]}

\journaltitle{Security and Privacy Protection for 5G-Enabled Internet of Things}
\articletype{Research Article} 

\begin{document}

\title{Privacy Preserving Data Analytics in 5G-Enabled IoT for the Financial Industry}

\author{%
		A. Cheng Lock LIM\addresslink{1},
    }
		
\address[1]{School of Computer Science and Engineering, Nanyang Technological University, Singapore}

\correspondence{Correspondence should be addressed to 
    	Cheng Lock LIM: limc0207@e.ntu.edu.sg}


\maketitle 

\begin{abstract}
Next-generation wireless networks like 5G promise faster speed, shorter latency, and the ability to connect more devices. Such benefits are set to make drastic changes to the future society, empowering smart cities, enabling autonomous cars, enhancing business processes, changing consumer behaviors, etc. In the financial industry, banks evaluate the deployment of Internet of Things (IoT) technologies and edge computing for better customer engagement, e.g., mobile branches on a vehicle, micro-ATM, self-service digital panel, etc. One of the trends is breaking down monolithic business application systems into micro-services for deployment on distributed edge servers, thus reducing network latency and improving services. Such movements pose challenges in protecting the security and privacy of business data between access points. This paper introduces a new architecture and protocol to tackle a use case for the financial industry. The solution assumes deploying a credit assessment model on an edge server. The model accepts and processes encrypted data submitted by potential customers seeking online credit assessments. The encrypted assessment results are sent back to the customers for decryption and interpretation. The data transmission rides on asynchronous communication, and the data protection uses Homomorphic Encryption. A proof-of-concept experiment shows that the proposed method can be achieved with a short response time and a reasonable prediction accuracy.    

\end{abstract}
    
\section{1. Introduction}
The emergence of next-generation cellular networks aims to provide higher bandwidth, faster speed, and shorter network latency. According to \cite{agrawal20165g}, the 5G network can support a low latency connection with a gigabit-per-second (Gbps) speed. This speed, coupled with a reduction in network latency, looks set to fuel creative inventions to change how businesses are conducted and how people live. For example, doctors controlling medical devices a thousand miles away from the patient's location can enable remote medical operations. Such an operation will require large data transmission and low network latency between the points to achieve an experience of "real-time."  

The new network also promises Massive Machine-Type communication, where the network will also be connecting up to one million devices within one square kilometer \cite{liu20165g}. This number far exceeds the number supported by the current 4G network. In other words, while 4G supports more mobile devices, the new network will support far more connections from other devices, including mobile devices and Internet of Things devices. The possibility of connecting so many devices also prompts many innovations.

In the financial industry, financial inclusion \cite{dev2006financial} refers to providing useful and affordable financial products and services to its target customer segments, such as individuals or businesses. Such financial products may include saving or current accounts to save money and perform financial transactions. More importantly, the products will also include personal loans or working capital loans, which are essential for individuals or businesses to fund their spending or start a small business. Hence, to achieve financial inclusion, the ability to reach and service the individuals or businesses would require connectivity, especially network or cellular connection. The connectivity will allow services to be conducted online and data to be transmitted over the wireless network. Under such an arrangement, the financial institution will have to design and deploy affordable financial services to serve the target segment with service quality, speed, and security. For security, protecting the data privacy of individuals and businesses will be essential. 

\subsection{1.1. Background}
One of the financial industry's trends is to re-evaluate the information technology systems' concentration in centralized data centers. Such concentration means existing, and new external network traffic will be transmitted to the data center and vice versa, possibly by traversing many network hops before reaching a destination. The hops will introduce network latency, and the sum of all latencies can also mean slower response time. As such, some organizations are evaluating re-architecting their traditional monolithic application system into micro-service architecture \cite{bucchiarone2018monolithic}. The latter aims to break down services into more modular and granular micro-services, of which each micro-service will carry out an atomic task and result in a shorter processing time. Deploying application systems on the edge servers will introduce a shift in security schemes compared to centralized system deployment in a data center. The data will traverse through the internet between different distributed edge nodes under an edge computing setup and require strong security protection for data transmission between them. 
 
Banks and finance companies are financial institutions providing financing (i.e., a loan) to needing customers. Before granting a loan, the Bank will assess the customer’s credibility in repaying a loan. Such assessment usually requires the customer to provide personal data (like age, occupation, year of employment, etc.) and financial data (income, debt, etc.). Such data can be deemed sensitive and can lead to privacy issues if such data is exposed. The Bank can collect such data directly using a hosted web page or through a bank’s mobile application in an online setting.

In some cases, the Bank could provide financing through business partners, and the credit assessment will be channeled through the partners. In such a case, the business partners may gain access to the customer's sensitive data, and the latter may raise concerns. Also, a bank might engage more than one business partner for one service, and any dispute on data privacy issues can lead to legal disputes, reputational or financial loss. Fortunately, \cite{chen2018logistic} provides a means to conduct a logistic regression analysis over encrypted data. In other words, a credit assessment estimation using logistic regression \cite{crook2007recent} can be conducted on ciphertexts, using Homomorphic Encryption (HE) scheme \cite{parmar2014survey}. Some years ago, the high computation cost of HE was a challenge in practice. However, performance improvement in HE schemes over the last few years has brought the possibility of practical HE implementation \cite{bensitel2016secure} \cite{ren2021privacy}. 
 
This paper introduces a new architecture and protocol that addresses the above scenario. An edge server, which the Bank owns, will evaluate the credit assessment on ciphertexts. The edge server will host a credit assessment model which is capable of performing an evaluation on ciphertexts. The ciphertexts are encrypted personal and financial data supplied by the customer. Homomorphic encryption (HE) scheme will protect the data and perform the evaluation on the ciphertexts. The HE capability is assumed to be embedded into the Bank’s edge server and a bank’s software development kit (SDK). The SDK can be embedded within the Bank's mobile application. Customer who wishes to request a credit assessment will install the Bank's mobile application on a personal mobile phone. The phone can also be seen as an Internet of Things (IoT) device on the new generation network. To cover the multi-parties set up, comprising Bank and business partners, an asynchronous publish/subscribe communication mechanism is deployed, similar to the typical IoT setting.  This mechanism is assumed to bridge a more effective communication between the customers and the Bank. In the experiment, a system is set up as per the above description and is tested with end-to-end data transmission. The test results, which measured the customer end's response time, show that the architecture and system design can support a privacy-preserving credit assessment over the above network setup.

The main contributions in this paper are summarised as follows: 
\begin{itemize}
    \item This paper introduces practical homomorphic encryption-based privacy-preserving data protection, which is generic in supporting many business use cases.
    \item This paper also introduces a practical protocol in using asynchronous IoT-related protocol to support business use cases.
    \item The experiment results show the usability and practicability of the proposed architecture and protocol.   
\end{itemize}

This paper's remaining sections are organized as: Section 2 covers the Related Works; Section 3 covers the Proposed Architecture, System Setup, and Proposed Protocol; Section 4 presents the experiments and discussions; Section 5 concludes this paper.

\section{2. Related Works}

\subsection{2.1. Logistic Regression (LR) Algorithm}

The purpose of a credit scoring model is to predict the capacity (equivalent to the financial obligation or creditworthiness) of a borrower to repay a loan. In this context, loan products cover unsecured lending such as credit lines or credit cards or secured lending such as auto loans or mortgage loans. The final prediction result for a loan application can be a binary value of zero or one. Such models allow financial institutions to assess a borrower before approving a loan adequately, thus minimizing the risk of a default and avoiding financial loss to the Bank. 

Logistic regression is commonly used for making a binary related prediction on the probability of default and the influence of different variables on a consumer's creditworthiness \cite{al2014credit}. The predictor variables can include occupation, annual income, outstanding debts, past credit history, etc., in our application. 

Consider the probability of default, $p = P(Y = 1)$ where its log-odds is assumed to be "default" and have a linear relationship with two predictor variables (denoted as $x_1, x_2$):

\begin{enumerate}
    \item The log-odds of $Y = 1$ can be written as a linear combination of the predictor variables:
    
    \begin{equation*}
    \label{eq:l1}
      l = log_{b} \frac{p}{1 - p} = \beta_{0} + \beta_{1}x_{1} + \beta_{2}x_{2}\\
    \end{equation*}

where $b$ is the base of the logarithm and $\beta_{i}, i=0,1,2$ are the weighting parameters of the model. $\beta_0$ is also known as the intercept or bias term.

    \item The odds can be computed by applying exponent to the log-odds:
    
    \begin{equation*}
    \label{eq:l2}
      \frac{p}{1 - p} = {b}^{\beta{0} + \beta_{1}x_{1} + \beta_{2}x_{2}} \\
    \end{equation*}
    
    \item The probability of $Y = 1$ can be derived by applying algebraic calculation and by dividing the terms by the exponent term:
    \begin{eqnarray}
    p &=& b^{\beta_0+\beta_1x_1+\beta_2x_2}(1-p) \nonumber \\
    (1+b^{\beta_0+\beta_1x_1+\beta_2x_2})p &=& b^{\beta_0+\beta_1x_1+\beta_2x_2} \nonumber \\
    p &=& \frac{b^{\beta_0+\beta_1x_1+\beta_2x_2}}{1+b^{\beta_0+\beta_1x_1+\beta_2x_2}} \nonumber \\
    p &=& \frac{1}{1 + {b}^{-(\beta_0+\beta_1x_1+\beta_2x_2)}} \nonumber \\
    p &=& {S_b({\beta_0 + \beta_1x_1 + \beta_2x_2})} \nonumber    
    \end{eqnarray}
    
    where $S_{b}$ is the sigmoid function with base $b$, and this function maps the predictions to a probability value between 0 and 1.
    
\end{enumerate}

The above formula shows the relationship between the parameters $\beta_{i}$, variables $x_{1}$, $x_{2}$, and the probability $P(Y=1)$. The parameters $\beta_{i}$ are usually obtained based on a training process utilizing a set of training data with a ground truth of the "default / not-default" label for each sample. 

In this work, the customer's input data, as described in Section 4.2, will form the features. The model training process will use a set of training data to compute the weights and the bias to meet the ultimate result, which is the target binary variable of "Loan Status." Then, a test is conducted on the learned model using a set of test data to measure the overall prediction accuracy.

\subsection{2.2. Homomorphic Encryption Scheme}

Homomorphic encryption (HE) is a type of encryption where the encrypted data's computation (or evaluation) can be done without decrypting the data. The evaluated result will also be an encrypted value, and the decrypted value is proved to be similar to the computing result based on the plain texts. This property enables Homomorphic encryption to be a good candidate for privacy-preserving storage and computation in an outsourcing setup. In other words, data can be encrypted by one party, and computation of the encrypted data can be performed by another party (i.e., under an outsourcing arrangement using commercial cloud computing).

It was until 2009 when Craig Gentry proposed the "bootstrapping" approach, and the first Fully Homomorphic Encryption (FHE) scheme \cite{gentry2009fully} was found. HE schemes started with homomorphic multiplication using the RSA scheme \cite{rivest1978data} and homomorphic addition utilizing the Paillier scheme \cite{paillier1999public}. These schemes with a single arithmetic operation are called Partially Homomorphic Encryption (PHE). Further development in this field has yielded another scheme called Somewhat Homomorphic Encryption (SWHE), where a limited number of arithmetic operations are allowed.     

This paper uses the Cheon-Kim-Kim-Song (CKKS)\cite{ckks} scheme to encrypt the data and perform an evaluation on the encrypted data. CKKS is an HE scheme that enables approximate arithmetic on real numbers, as compared with other HE schemes such as BGV\cite{brakerski2014leveled} or BFV\cite{bfv} which are working on integers only. CKKS scheme works on polynomial and allows addition, multiplication, rescaling, and rotation to be applied to polynomials. The message $m$ is always the real value to be encrypted, and $m$ (usually in a vector form) will be encoded into a plaintext $pt(X)$ represented in polynomials. The encoded message $pt(X)$ can then be encrypted to form the ciphertext $ct$, and $ct$ will store the significand part of the numbers and the scaling vector. The CKKS scheme also provides relinearization and rescaling functions to reduce the noise after every HE operation, especially multiplicative operations. The evaluated result $\Bar{m}$ is recovered by first decrypting the ciphertexts to get the plaintext, followed by decoding the plaintext (includes division by a scaling vector).

This paper does not intend to explain CKKS in detail, and the rest of the paper will focus on using CKKS for a use case instead. Hence, the high-level CKKS implementation is summarized as follows:
\begin{itemize}
    \item \textbf{Key generation:} $(pk,sk) \leftarrow$ CKKS.Keygen$(pp)$. Takes the public parameter \textit{pp} as an input and returns a pair of public key $pk$ and secret key $sk$.
    \item \textbf{Encode:} $pt(X) \leftarrow$ CKKS.Encode$(m)$. Encode the clear text $m$ into polynomial plain-text $pt(X)$ through polynomial integer ring $\frac{\mathbb{Z}[X]}{(X^{N}+1)}$. 
    \item \textbf{Encrypt:} $ct \leftarrow$ CKKS.Encrypt$(p(X), pk)$. Encrypts the polynomial plain-text using the public key $pk$ and outputs a ciphertext $ct \in \{0,1\}^{*} $.
    \item \textbf{Evaluate:} $(\Bar{ct}) \leftarrow$ CKKS.Evaluate$(ct, pk ,\mathcal{C})$. Given a ciphertext $ct$, the public key $pk$ and circuit $\mathcal{C}$, output a new ciphertext $\Bar{ct})$. Relinearisation and re-scaling are usually performed after the HE computation.
    \item \textbf{Decrypt:} $pt(\Bar{X}) \leftarrow$ CKKS.Decrypt$(\Bar{ct},sk)$. Decrypts the ciphertext into a polynomial plaintext using the secret key.
    \item \textbf{Decode:} $\Bar{m} \leftarrow$ CKKS.Decode$(pt(\Bar{X}))$. Returns the clear text message $\Bar{m}$. 
\end{itemize}

One drawback in using CKKS is the higher accuracy reduction due to the noise added during encryption and approximation arithmetic. However, this problem is not apparent in the experiment because the use case is Logistic Regression and did not involve many multiplications.

\subsection{2.3. Asynchronous Publish / Subscribe}

Asynchronous mode of communication is suggested in this paper's use case for a few reasons. In a high-density network environment where devices compete for connectivity, synchronous communication will hold up connectivity if there are prolonged responses from layers above the network, e.g., application layers. Also, while the existing 4G network merely serves mobile devices, this may not be the case in the new generation network. There will be no clear distinction between mobile devices and Internet Of Things (IoT) devices in the new generation network. Hence, traditional online services and applications will have to evaluate if current synchronous communication (e.g., HTTP) should only be reserved for mission-critical services. In this paper, the use case is a customer-initiated online credit assessment request. Such a request is assumed to be a non-time critical service. Hence, the paper takes the opportunity to introduce asynchronous communication and conduct an experiment to demonstrate asynchronous communication that serves the intended use case and provides data privacy protection.  

One of the most popular IoT protocols is Message Queue Telemetry Transport (MQTT) which implements lightweight publish/subscribe services. There has been development in enabling enhanced security on MQTT. In \cite{prantl2021performance}, Transport Layer Security (TLS) can only protect network traffic between publisher and broker or subscriber with the broker. The TLS cannot guarantee end-to-end encryption between publishers and subscribers. A Secure MQTT was proposed in \cite{singh2015secure}, and it enables security using lightweight Attribute-Based Encryption (ABE). These types of enhanced security would not be suitable for data protection in our use case because end-to-end data protection is required in our use case. As a result, application-level data encryption with homomorphic encryption is selected. 

\section{3. Proposed Architecture and System Setup}

This section covers the proposed architecture and system setup. 

\subsection{3.1 Architecture}
The architecture comprises four layers: message sender, message broker, message receiver, and a server application.  
 \begin{figure*}[ht]
    \centering
    \includegraphics[width=16cm]{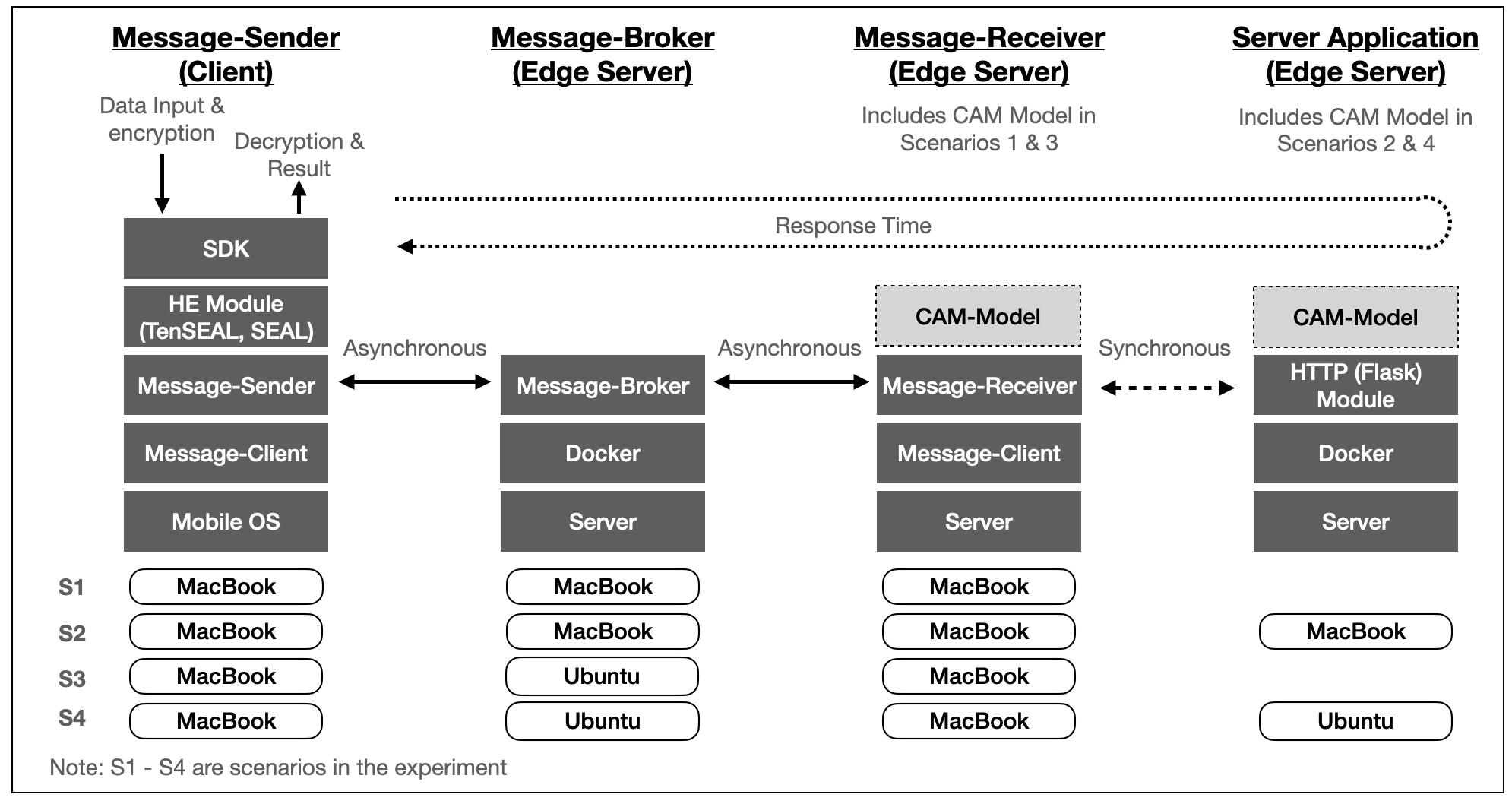}
    \caption{Overview of the Round-trip Setup}
    \label{fig:Figure1}
\end{figure*}
 
\begin{itemize}
  \item \textbf{Message Sender (MS)}. This sender is the initiator of a credit assessment request. The message requester is a message client to the Publish/Subscribe service. The message client is assumed to be a client module embedded within a mobile application in this setup. The mobile application will manage the HE encryption context, capture the data input, encrypt the data using HE, output the ciphertexts to the message client. Then, the message client will send the ciphertexts to the message broker and wait for a response. The experiment will record round-trip response times and selected processing times from sending a message request to receiving a message response in this setup. 
  \item \textbf{Message Broker (MB)}. This broker is a message broker server that manages the asynchronous publish/subscribe service. The broker manages "topics" (which can be viewed as "Channels") and relays messages between the publisher and subscribers of each topic. In other words, it bridges message clients at both sides of the messaging and delivers individual requests and responses to the designated message clients.
  \item \textbf{Message Receiver (MR)}. This receiver is the "server leg" of the publish/subscribe service. The receiver listens to a topic, receives the message, and posts or processes the credit assessment message (depending on the experiment scenarios shown in Figure 1). The receiver will return the broker's response message after receiving the encrypted result from the credit assessment. Although the MR is also a message client to the publish/subscribe service, its main job is to relay the message request to the credit assessment model and vice versa.
  \item \textbf{Server Application}. This server application is a microservice application that is assumed to run on an edge server. The application runs within a docker container and is installed with a Flask application (which enables an HTTP service). The server application receives and processes a credit assessment request via an HTTP request. This server application will host the logistic regression model in the setup, and the model will produce an encrypted credit assessment score from the requester's ciphertexts. The server application will return an HTTP Response with the encrypted credit score. This HTTP module is omitted in two experiment scenarios, and the credit assessment module is hosted with the Message Receiver.
\end{itemize}

\subsection{3.2 Setup}
 
It is assumed that the Bank must set up four components. They are :

\begin{enumerate}
  \item \textbf{Credit Assessment Model (CAM)}. It is assumed that the Bank will install a pre-trained credit assessment model at its edge server, either on the Message Receiver server or the Server Application server. The model is deployed to these edge servers through standard system deployment procedures. It is assumed that the edge server is secured with standard security defense and that the CAM and the model parameters are not required to be encrypted.
  
  \item \textbf{Homomorphic Encryption module (HE)}. This module is assumed to have two components. First, HE capability is delivered within a Software Development Kit (SDK) embedded in a mobile application. Next, the HE module also requires a server-side module to perform the computation on ciphertexts. The SDK will implement a Homomorphic Encryption (SDK-HE) module and a message client. The SDK-HE module will implement the CKKS library and perform cryptographic key generation, secure data input, data encoding and decoding, and data encryption and decryption. The SDK should also contain the mean and standard deviation values (to be obtained during the logistic regression model training process) to normalize input data before encrypting the input data. The SDK will output ciphertexts that represent the credit assessment request. The message client will then send the ciphertexts to the message broker. Upon receiving the message broker's message, the SDK also decrypts and decodes the credit assessment result. The SDK can be embedded into the Bank's mobile application, and it can store the cryptographic keys securely using the mobile device's secure elements. The SDK library can apply obfuscation to prevent any reverse-engineering to uncover the program codes. 

The server-side will also implement a Homomorphic Encryption (S-HE) module that will perform the logistic regression computation on the ciphertext.   
  
  \item \textbf{Message Receiver (MR)}. As described in Section 3.1, the message receiver is the entity between the message broker and the CAM module. The MR listens to the topic's queue on the message broker, registers the message, posts the message to the edge server, receives a response from the server application, and returns the response to the message broker. It is assumed that the Bank manages the MR (applicable to all four scenarios in the experiment). There can be multiple MRs to meet the request volume in the proposed setup, and each MR should have a designated CAM server close to it. 
  
  \item \textbf{Message Broker (MB)} The Bank can shortlist suitable partners or service providers who operate the publish/subscribe message broker service on behalf of the Bank. The setup at the service provider will be:

\begin{enumerate}
    \item Create a topic queue for the Bank's credit assessment requests.
    \item Receive incoming requests from message clients.
    \item Relay the requests to a subscriber (Message Receiver).
    \item Process returning (response) requests from Message Receiver.
    \item Relay the individual responses to the respective message clients.
\end{enumerate}

\end{enumerate}
It is worth noting that the credit assessment data and results will remain encrypted on the round-trip communication from the SDK to the message clients, the message broker, the message receiver, and the CAM server. It is also assumed that the Bank will also implement proper security authentication and authorization schemes between the online services. 

\subsection{3.3 Proposed Protocol}
Protocol 1 presents the proposed protocol for the Credit Assessment Scheme with Homomorphic Encryption.

\begin{table}[h!]
\begin{tabularx}{\linewidth}{X}
\hline
\textbf{Protocol 1.} {Credit Assessment Scheme with Homomorphic Encryption} \\
\hline
\textit{Inputs.} All message-senders are denoted with  $MS_i$ and each $MS$ possesses a set of data elements $m = x_1,...,x_n$ for a credit assessment.
\sbline 
\textit{The protocol:}
\begin{enumerate}
  \item \textbf{Setup.} Through the embedded HE module, each $MS_i$ creates a set of CKKS security contexts $SC_{i}$, comprising CKKS security parameters (as shown in Table 1), and performs a key generation to produce a secret key $sk_i$ and a public key $pk_i$. 
  \item \textbf{Process.}
  \begin{enumerate}
    \item Each $MS$ encodes the message $m$ into a polynomial plaintext $pt(m)$.
    \item The $MS$ encrypts $pt(m)$ with $pk$, thus producing a ciphertext $ct = e_{pk}(pt(m))$, and a set of public security contexts $pSC$.   
    \item Each $MS$ sends $ct$ and $pSC$ to the $MB$.
    \item The $MB$ relays $ct$ and $pSC$ to $MR$.
    \item The $MR$ delivers $ct$ and $pSC$ to the $CAM$ module.
    \item The $CAM$ module uses $pSC$ to run a logistic regression $LR$ evaluation on $ct$ thus producing a result $\bar{ct} = LR(ct))$, where $\bar{ct}$ is also encrypted.
     \item The encrypted result $\bar{ct}$ is then relayed back to $MR$, to $MB$ and to the $MS$.
     \item The $MS$ decrypts $d_{sk}(\bar{ct}) = d_{sk}(LR(ct)) = d_{sk}(LR(e_{pk}(x_i,...,x_n))) = LR(x_i,...,x_n) \approx pt(\bar{X})$, where $pt(\bar{X})$ is the encoded value of the LR result.
     \item The $MS$ decodes $pt(\bar{X})$ to produce $\bar{m}_i$, which is the approximate result in cleartext from the $LR$, based on the original data of $x_i,...,x_n$.
     \item $\bar{m}$ can be fed into a sigmoid function $S_b(X)$ to derive a binary output.
    \end{enumerate} 
\end{enumerate}
\\ \hline
\end{tabularx}
\end{table}

\section{4. Experiments and Discussions}

\subsection{4.1 Experiment Setup}
The objective of the experiment is to measure the round-trip response time in a simulated 5G environment. The measurement will evaluate the feasibility of setting up such an architecture over a 5G network, with communication using IoT protocol and computing nodes running as edge servers. The simulated environment is set up on two machines, and they are connected to a WiFi access point. These two machines can be identified as "MacBook" and "Ubuntu" in Figure 1. The first machine is a MacBook Pro (13-Inch, 2020 Model with 2 GHz Quad-Core Intel Core i5, 16GB RAM). This machine runs the full setup for scenario 1 and scenario 2. The machine also simulates the Message-Receiver in Scenario 3. The second machine is a Linux Ubuntu v18.04 Workstation running on Intel(R) Core(TM) i3-4170 CPU @ 3.70GHz, 32GB memory. The machine hosts the Message Broker and the bank server application for Scenario 3 and hosts the Message-Receiver (with CAM) in Scenario 4. Thus, the message communications are routed back and forth, mimicking as close to a real and physical setup. Both the Message Broker and Bank's Server application are programmed using Python, and both services run on docker, simulating that both services are running on edge servers. The Server application also runs Flask for the HTTP service. No transport layer security is enabled in the experiment.

The data transmission between Message Senders, Message Receiver, and Message Broker uses the asynchronous publish/subscribe service of  Message Queuing Telemetry Transport (MQTT). In the test scenario where there is a CAM server, the Message Client will send the CAM server's request using Hypertext Transfer Protocol (HTTP). Between the services, data are transmitted using Javascript Object Notation (JSON).
 
The data for credit assessment is input by the user and fed to the SDK. The SDK is assumed to validate the data before applying data normalization and encryption. 
 
The Homomorphic Encryption (HE) module uses TenSEAL \cite{TenSEAL}, an MIT-licensed software package developed on Python. TenSEAL's underlying HE package is Microsoft SEAL \cite{sealcrypto}. According to Microsoft SEAL online documentations, SEAL is supported on Android and iOS platforms, which would assume that the HE module can be embedded in a mobile application to be used by the request originator. The TenSEAL package requires initializing a security context that will determine the encryption's security strength. The following security context parameters (Table 1) for 128-bit security are used in the experiment.
 
\begin{table}[ht]
\caption{Security Context Parameters}
\begin{tabular}{ |p{1.5cm}|p{1.5cm}|p{1.5cm}|p{1.5cm}|  }
 \hline
  Polynomial modulus degree & Coefficient modules bits sizes & Global Scale & Galois key\\
  \hline
  4096 & 40, 20, 40 & ${2^{20}}$ & Generated\\
  \hline
\end{tabular}
\label{Tab:Tcr}
\end{table}
 
One secret key and one public key are required to enable computation on ciphertexts at the Server application. The security context is required to be made public but without the secret keys. From the experiment, the security context is a file of 6 Megabytes where the Galois key forms the bulk of the memory sizes, according to the SEAL design. 
 
The public security context and the encrypted data are serialized into Bytes object for data transmission. Both data in the Bytes object will undergo Base64 encoding before they are transmitted before points. 

\subsection{4.2. Dataset} This paper's use case is for potential new customers to request online credit assessments with a Bank. Each customer is expected to submit a required set of personal data to the Bank. The Bank will make a credit assessment based on the data and return an assessment result to the customer. The result is typically a numeric score, indicating the chance for the customer to obtain a loan from the Bank. Instead of showing the numeric value directly to the customer, it is assumed that the Bank will wrap the numeric value with customized text messages. The messages will be better in conveying the assessment result and managing the customer's expectations. 

The dataset used in this research is the Credit Risk Dataset obtained from Kaggle \cite{CreditRiskDataset}. The data contains more than 32,000 consumer lending records. The data has 11 feature points: annual income, home status, age, employment length, loan intent, loan amount, loan grade, interest rate, loan-to-income ratio, historical default, and credit history length. One other variable, "loan status," is a binary variable that indicates default status (0 is non-default, 1 is the default). The loan status is used as the target variable for training a logistic regression model. Records with missing values are dropped, and outliers are also dropped. Feature points that are non-numerical are converted to numeric values. A logistic regression model is trained with the resultant dataset.

\subsection{4.3 Test Scenarios}

The test scenarios are:

\begin{enumerate}
    \item Train a credit assessment model (CAM) using the dataset (specified in Section 4.2). During this training, the dataset is converted from 11 feature points to 26 feature points due to certain categorical variables being converted into indicator variables. The model is trained using the PyTorch library \cite{NEURIPS2019_9015}, and the model prediction accuracy on plaintext data is 80.7\%. 
    \item Set up one CAM server (edge server) running Python-Flask on docker and deploy the server application's CAM model. The same CAM model is also deployed at the Message Receiver for scenarios 1 and 3. 
    \item Set up a Message Broker service, running docker and Mosquitto-MQTT broker \cite{mosquittoMQTT}.
    \item Set up one Message Receiver on an edge server, using Paho-MQTT client library \cite{mosquittoMQTT}.
    \item Set up five Message Senders, each sending 20 concurrent requests to the message broker and CAM Server simultaneously. Each Message Sender is labeled with a unique identifier, e.g., Message\_Sender\_1 to Message\_Sender\_5. Each message is also tagged with a Co-relation ID, which is specified by the MQTT protocol. One set of security contexts is created for each Message Sender client. The security context is re-used to encrypt and decrypt the data for the 20 requests sent by each client. The Message Sender also implements the Paho-MQTT client library.
\end{enumerate}

\begin{figure}[ht]
    \centering
    \includegraphics[width=8cm]{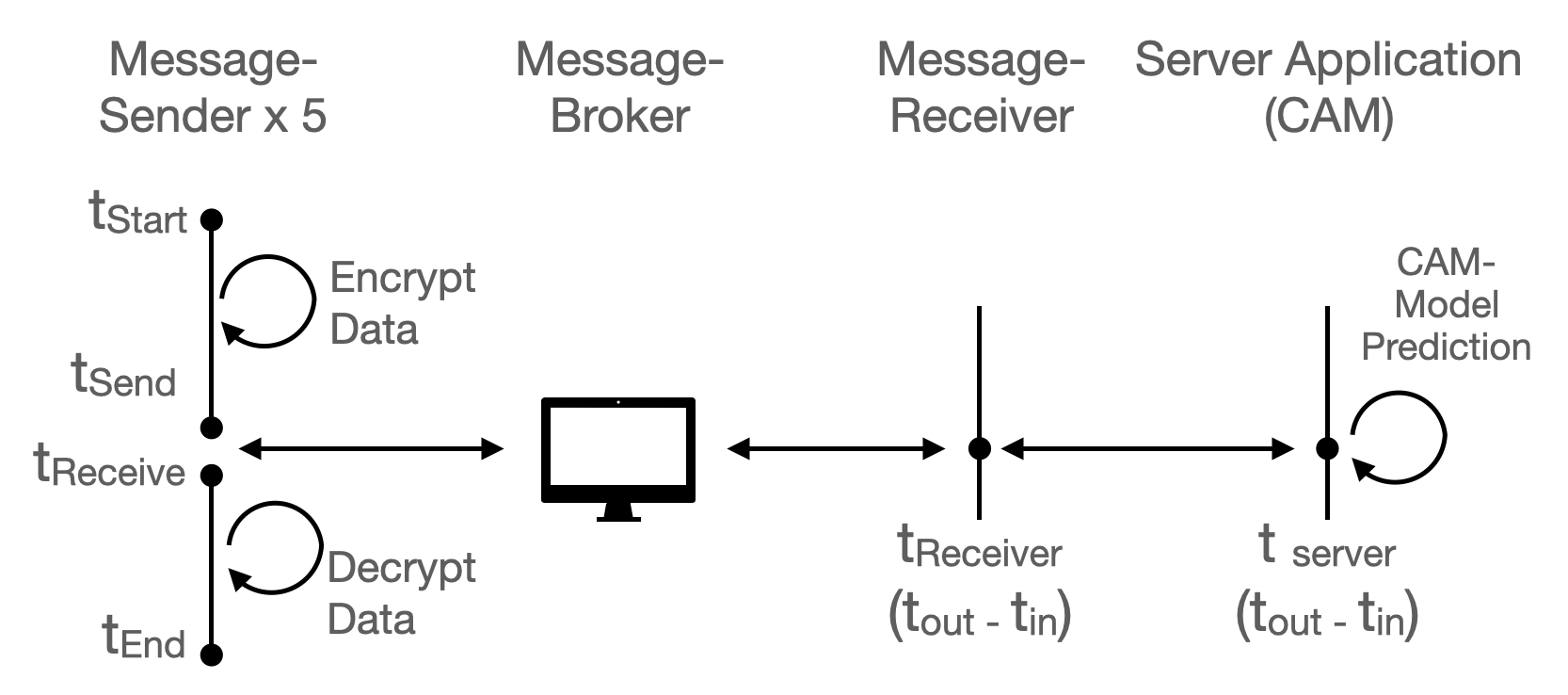}
    \caption{Overview of time recording}
    \label{fig:Figure2}
\end{figure}

\subsection{4.4 Time Recording}

Recording of the timestamp is carried out as per Figure 2. Each timestamp record is tagged with the Message Sender ID and its respective Co-relation ID.

\begin{table*}[t]
  \centering
  \caption{Experimental results (ms=milli-second)}
  \begin{tabular}{ |p{2.4cm}|p{1.7cm}||p{1.6cm}|p{1.5cm}|p{1.6cm}||p{1.6cm}|p{1.5cm}||p{1.6cm}|}
 \hline
  Scenario & Round-trip (Start-End) (ms) & Start-Send (ms) & Send-Receive (ms) & Receive-End (ms) & Message-Receiver (ms) & Server Application (ms) & Prediction (\%)\\
  \hline
  Scenario 1 - 3 layers on MacBook, No HTTP Server & 950 & 272 & 675 & 1.39 & 115 & -- & 73\% \\
 \hline
  Scenario 2 - 4 layers on MacBook & 1,645 & 312 & 1,330 & 1.97 & 498 & 419 & 81\% \\
  \hline
  Scenario 3 - 2 layers on  MacBook, Broker on Server, No HTTP Server & 3,266 & 265 & 2,998 & 2.65 & 111 & -- & 75\%\\
  \hline
  Scenario 4 - 2 layers on  MacBook, Broker \& HTTP on server & 5,177 & 275 & 4,898 & 2.96 & 1,474 & 1,209 & 75\%\\
  \hline
  \hline
  Additional Test on Scenario 2 with higher polynomial modulus degree & 9,787 & 3,726 & 6,036 & 23.55 & 3,358 & 2,285 & 73\%\\
  \hline
  \end{tabular}
  \label{tab:1}
\end{table*}

The Message Sender program consumes a file containing 12,404 test data, and each row contains 27 feature points. Twenty-six points are the data, and one data point is the ground truth. The program creates a Pandas data frame object for each test client. The program randomly picks a row of test data from the dataset and registers the ground truth with a unique identifier and timestamp in the data frame. 
 
The Message Sender program logs the timestamps and time records in the data frame upon receiving the response message. The time records include $t_{receiver}$ and $t_{server}$, which are embedded in the response message. The encrypted result is decrypted and recorded in the data frame. The result is tallied with the ground truth to validate the accuracy of the encrypted computation. Various timestamps will be computed to validate the response time in such a setting.  

\subsection{4.5 Test Results}

The result in Table-2 can be summarized as follows:

\begin{itemize}
    \item With 128-bit security and 26 feature points, the time spent on encryption (start-send) and decryption (receive-end) is consistent over the scenarios. The average sum of these two timings is 282 ms, less than 8\% of the round-trip time for scenario 3 and scenario 4, respectively.
    \item The longer round-trip response times for scenario 3 and scenario 4 show the effect of network latency since the Message Broker is hosted on a separate machine.
    \item The inclusion of an HTTP server in scenario 4 increases its round-trip time by 2 seconds more than scenario 3. In other words, the Message Receiver node can host the CAM model to save the data transmission.
    \item Prediction results from the four scenarios, based on ciphertexts, yielded an average of 76\%.
    \item With 128-bit security and 26 feature points, the average size of payload for data transmission is 8.3 Megabytes, which is optimal for this HE scheme.
    \item Overall, the response times for scenario 3 (i.e., 3.26 seconds) should be acceptable for the use case of a quick online credit assessment request based on asynchronous communication. 
\end{itemize}

It is worth noting that the Message-Receiver timing includes that of the HTTP server. Their time difference is small, and that is the time spent to route the HTTP response to the message broker. Another point to note is that very high accuracy on the prediction is not the focus of this experiment. Instead, this experiment focuses on implementing homomorphic encryption for privacy preservation over edge servers and asynchronous communication, which mimic a new network setup.

\subsection{4.6 Discussions}

In the proposed protocol, the cleartext input data does not leave the SDK. And the computation at the server side is performed on the ciphertext. The secret key and public key are also generated and stored within the SDK. It is assumed that the SDK will leverage a secure element in the mobile phone to store the cryptographic keys securely. The cryptographic keys can be discarded when the resultant ciphertext is decrypted and the user is satisfied with the completion of the request. A new set of cryptographic keys can be generated when a new credit assessment request is initiated.

There is a scenario where an honest Message Broker (a third party to which the Bank outsources the MB service) turns into an honest but curious party who wants to obtain any sensitive data. The security assurance from this proposed protocol is the security context's encryption strength. The level of data sensitivity will determine the encryption strength. However, it is worth noting that the higher the security strength, the larger the payload and the time required for the data transmission. The last row of Table 2 is an additional Scenario-2 test conducted with a higher security context of polynomial modulus degree of 8,192, with coefficient modulus bit sizes = [40, 21, 21, 21, 21, 21, 40]. The payload is increased to 128 Megabytes for the same experimental data. This size is 16 times that of the payload with a polynomial modulus degree of 4,096. The size is mainly due to the increase in Galois key and Relinearization key being exported to the public security contexts. This increase in size is expected of the SEAL library \cite{sealcrypto}. As a result, the round-trip performance is also lengthened to 9,787 milliseconds, six times that of Scenario 2.

Despite being hard to crack as compared to traditional prime-based cryptographic schemes, a new exploit has been found in CKKS\cite{exploit}. The exploit allows attackers to completely recover the secret key with high probability and using modest running time. According to \cite{cheon2020remark}, different libraries which implemented the CKKS scheme have taken steps to mitigate the libraries against the attack. At a high level, such mitigations adopt a similar practice of affixing a small error after the decryption step to avert dependency on the secret key and encryption randomness. Libraries such as HEANN, HELib, and PALISADE, have enhanced their libraries while SEAL advised the secret key owner to safe keep the SEAL ciphertexts and decryption results as private information. Subsequent implementation of this protocol should consider the development of each HE library.

The experiment results from this paper showed that Banks could leverage the proposed architecture and protocol to enlarge their financial inclusion offering to a larger segment of the needy population while preserving data privacy. On the other hand, it is also possible to adopt a Multi-Key FHE \cite{chen2019efficient} solution where the Bank's model parameters are protected via encryption on the edge server. The Multi-Key FHE solution will involve more communication rounds because it will have to be partially decrypted by the model owner. The decrypted result must be communicated from the model owner to the requester for the latter to decrypt the computation result fully.

\section{5. Conclusion}

Banks evaluate the deployment of Internet of Things (IoT) technologies and edge computing for better customer engagement in the financial industry. One of the trends is breaking down monolithic business application systems into micro-services for deployment on distributed edge servers, thus reducing network latency and improving services. Such movements pose challenges in protecting the security and privacy of business data between access points. 

This paper introduces a new architecture and protocol for the financial industry to tackle a use case, credit assessment modeling. The approach proposes to deploy a credit assessment model on an edge server that accepts and processes homomorphically encrypted data submitted by potential customers seeking online credit assessments. The encrypted assessment results are sent back to the customers for decryption and interpretation.

The proposed approach has been implemented by adopting asynchronous communication, edge computing, and homomorphic encryption to evaluate the response time and prediction accuracy. Experiment results proved that it is practical for organizations to enhance existing application services while preserving data privacy and improving response time.  

\printbibliography

\end{document}